\documentclass{pasj00}

\begin{document}
\SetRunningHead{T.Nagayama et al.}{Astrometry of G48.61+0.02 with VERA}

\title{Astrometry of Galactic Star-Forming Region G48.61+0.02 with VERA}

\author{Takumi    \textsc{Nagayama},\altaffilmark{1}
         Toshihiro \textsc{Omodaka},\altaffilmark{2}
         Toshihiro \textsc{Handa},\altaffilmark{3} 
         Mareki    \textsc{Honma},\altaffilmark{1} \\
         Hideyuki  \textsc{Kobayashi},\altaffilmark{1}
         Noriyuki  \textsc{Kawaguchi},\altaffilmark{1}
         and
         Yuji \textsc{Ueno}\altaffilmark{1}
         }
\altaffiltext{1}{Mizusawa VLBI Observatory, National Astronomical Observatory of Japan, \\
                   2-21-1 Osawa, Mitaka, Tokyo 181-8588}
\email{takumi.nagayama@nao.ac.jp}
\altaffiltext{2}{Graduate School of Science and Engineering, Kagoshima University,\\
                   1-21-35 K\^orimoto, Kagoshima, Kagoshima 890-0065}
\altaffiltext{3}{Institute of Astronomy, The Universe of Tokyo, 2-21-1 Osawa, Mitaka, Tokyo 181-0015}

%

\KeyWords{Galaxy: kinematics and dynamics --- stars: individual (G48.61+0.02)} 

\maketitle

\begin{abstract}
We performed the astrometry of H$_2$O masers 
in the Galactic star-forming region G48.61+0.02
with the VLBI Exploration of Radio Astrometry (VERA).
We derived a trigonometric parallax of $199 \pm 7$ $\mu$as,
which corresponds to a distance of $5.03 \pm 0.19$ kpc.
The distance to G48.61+0.02 is about a half of its far
kinematic distance, which was often assumed previously.
This distance places G48.61+0.02 in the Sagittarius-Carina arm and near
the active star forming region and the supernova remnant W51.
We also obtained the three dimensional motion of G48.61+0.02, and
found that it has a large peculiar motion of $40\pm5$ km s$^{-1}$.
This peculiar motion would be originated with 
the multiple supernovae explosions in W51,
or the streaming motion across the Sagittarius-Carina arm.
\end{abstract}


\section{Introduction}

It is important to measure the distance accurately,
because most physical properties of 
individual sources, such as sizes, masses, luminosities
depend critically on the distance.
The kinematic distance,
which is derived from the observed radial velocity and the Galactic rotation model,
is widely used to estimate the distance from Sun to the source 
in the Milky Way Galaxy (MWG).
However, the kinematic distance would be different from the real distance,
in the case that the source motion is affected by 
the local activities such as the supernova (SN) and the super-bubble,
the potential of Galactic bar, and the streaming motion of the spiral arm.
Recent Very Long Baseline Interferometry (VLBI) astrometric observations
determined the distances of the numerous maser sources directly
from the parallax measurements.
Some of VLBI observations report that 
there is a difference between the parallactic distance and the kinematic distance
(\cite{sat08}; \cite{bru09}; \cite{san09}).
This is because that these observed sources have large peculiar motions 
possibly originated with the expanding super-bubble (\cite{sat08}), 
the gravitational perturbations from the Galactic bar (\cite{bru09}), and
the expanding 3 kpc arm (\cite{san09}).

G48.61+0.02 is a massive star-forming region 
located within \timeform{1D} (100 pc at the distance of 5 kpc)
from W51, which is one of the most energetic sources in the MWG.
However, the local standard of rest (LSR) velocities of two sources 
are different by approximately 40 km s$^{-1}$.
The LSR velocities of G48.61+0.02 and W51 
are $v_{\rm LSR} = 19\pm1$ km s$^{-1}$ (\cite{bro96}; \cite{lop09}; \cite{cod10})
and $58 \pm 4$ km s$^{-1}$ (\cite{sat10}), respectively.
Therefore, their kinematic distances are also different.
The near and far kinematic distances of G48.61+0.02 are 
derived to be 1.4 and 9.8 kpc, respectively,
in the case that IAU recommended values of the Galactic constants, 
$R_0 = 8.5$ kpc and $\Theta_0 = 220$ km s$^{-1}$, and the flat rotation are assumed.
The 21 cm wavelength HI absorption measurements suggest that
G48.61+0.02 is located at the far kinematic distance (\cite{sat73}).
The parallactic distance of W51 is measured to be $5.41^{+0.31}_{-0.28}$ kpc 
by the H$_2$O maser astrometric observations 
with the Very Long Baseline Interferometry (VLBA) (\cite{sat10}), 
and it is consistent with the kinematic distance.
Therefore, it is suggested that 
G48.61+0.02 is far away from W51 in the line of sight direction,
{and} these two sources are not physically associated (\cite{sat73}).
However, W51 is composed of the active sources
such as two complex H\emissiontype{II} regions W51 A and W51 B, 
and the SN remnant W51 C (\cite{koo02}).
It would be possible that
G48.61+0.02 has a large peculiar motion affected by their activities.
We can measure the distance of G48.61+0.02 directly by the parallax,
since the H$_2$O masers are associated with G48.61+0.02 (\cite{kur05}).
In order to determine the accurate distance and the reliable
location of G48.61+0.02 in the MWG,
we conducted multi-epoch phase referencing observations
of the H$_2$O masers with VERA.


\section{Observations and Data Reductions}

We observed H$_2$O masers in the Galactic star-forming region 
G48.61+0.02 with VERA at 12 epochs from 2005 January to 2006 March.
We present the data of 8 epochs that were obtained
with full 4-station array (Mizusawa, Iriki, Ogasawara, and Ishigaki-jima)
under relatively good conditions.
The epochs are as follows (day of year): 024, 052, 141, 299, 334, 341 in 2005,
and 050, 086 in 2006.
During about 10 hours at each epoch,
the H$_2$O $6_{16}$-$5_{23}$ maser at 
a rest frequency of 22.235080 GHz in G48.61+0.02 and 
two background sources, J1922+1530 and J1924+1540 
were simultaneously observed using the dual-beam system of VERA.
The typical flux densities of J1922+1530 and J1924+1540 are
240 and 480 mJy, respectively.
The separation angles between 
G48.61+0.02 and two background sources, J1922+1530 and J1924+1540,
are \timeform{1.66D} and \timeform{2.02D}, respectively.
During the observations, 
the instrumental phase difference between the two beams was
measured continuously by injecting artificial noise sources 
into both beams (\cite{hon08a}).
The left-hand circular polarizations were recorded onto magnetic tapes
at a rate of 1024 Mbps with 2-bit quantization,
providing a total bandwidth of 256 MHz, which consists of 16 of 16 MHz IF sub-band.
The filtering of IF-sub-bands was made using the VERA digital filter (\cite{igu05}).
One IF sub-band was assigned to G48.61+0.02, and the other 15 IF
sub-bands were assigned to position reference sources, respectively.
Correlation processing was carried out on the Mitaka FX correlator.
The frequency and velocity resolutions for G48.61+0.02
were 15.625 kHz and 0.21 km s$^{-1}$, respectively.

Data reduction was conducted using 
the NRAO Astronomical Image Processing System (AIPS).
The amplitude calibration was performed using 
the system noise temperatures logged during the observations
and the antenna gains.
For phase-referencing, a fringe fitting was
made using the AIPS task FRING on J1922+1530 and J1924+1540
with a typical integration time of 1 min and a time interval of 30 sec.
The solutions of the fringe phases, group delays, and delay rates
were applied to G48.61+0.02 in order to calibrate the visibility data.
Phase and amplitude solutions obtained from self-calibration of background sources
were also applied to G48.61+0.02.
Visibility phase errors caused by the Earth's atmosphere were calibrated
based on GPS measurements of the atmospheric zenith delay, 
which is mostly due to tropospheric water vapor (\cite{hon08b}).
We made spectral-line image cubes using the AIPS task IMAGR around masers with
$1024\times1024$ pixels of size 0.05 milliarcsecond (mas) after the calibration.
The typical size of the synthesized beam was $1.3\times0.9$ mas 
with position angle of \timeform{-40D}.
The rms noises for each images were approximately 0.1--1 Jy beam$^{-1}$.
The signal-to-noise ratio of 7 was adopted as the detection criterion.


\section{Results}

\begin{figure}
  \begin{center}
    \FigureFile(80mm,80mm){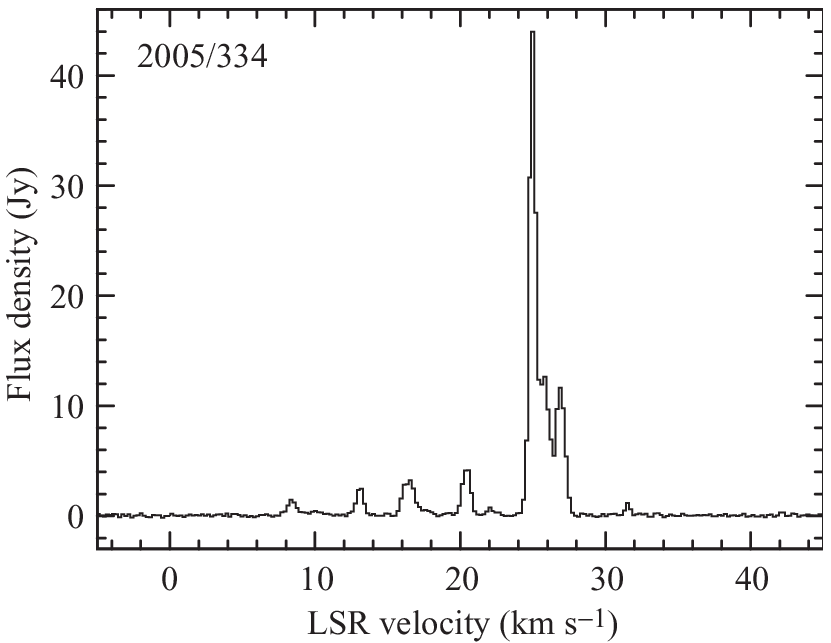}
  \end{center}
  \caption{A cross-power spectrum of the G48.61+0.02 H$_2$O masers 
            obtained with Mizusawa-Iriki baseline in 2005/334.}
  \label{fig:1}
\end{figure}

\begin{figure*}
  \begin{center}
    \FigureFile(160mm,90mm){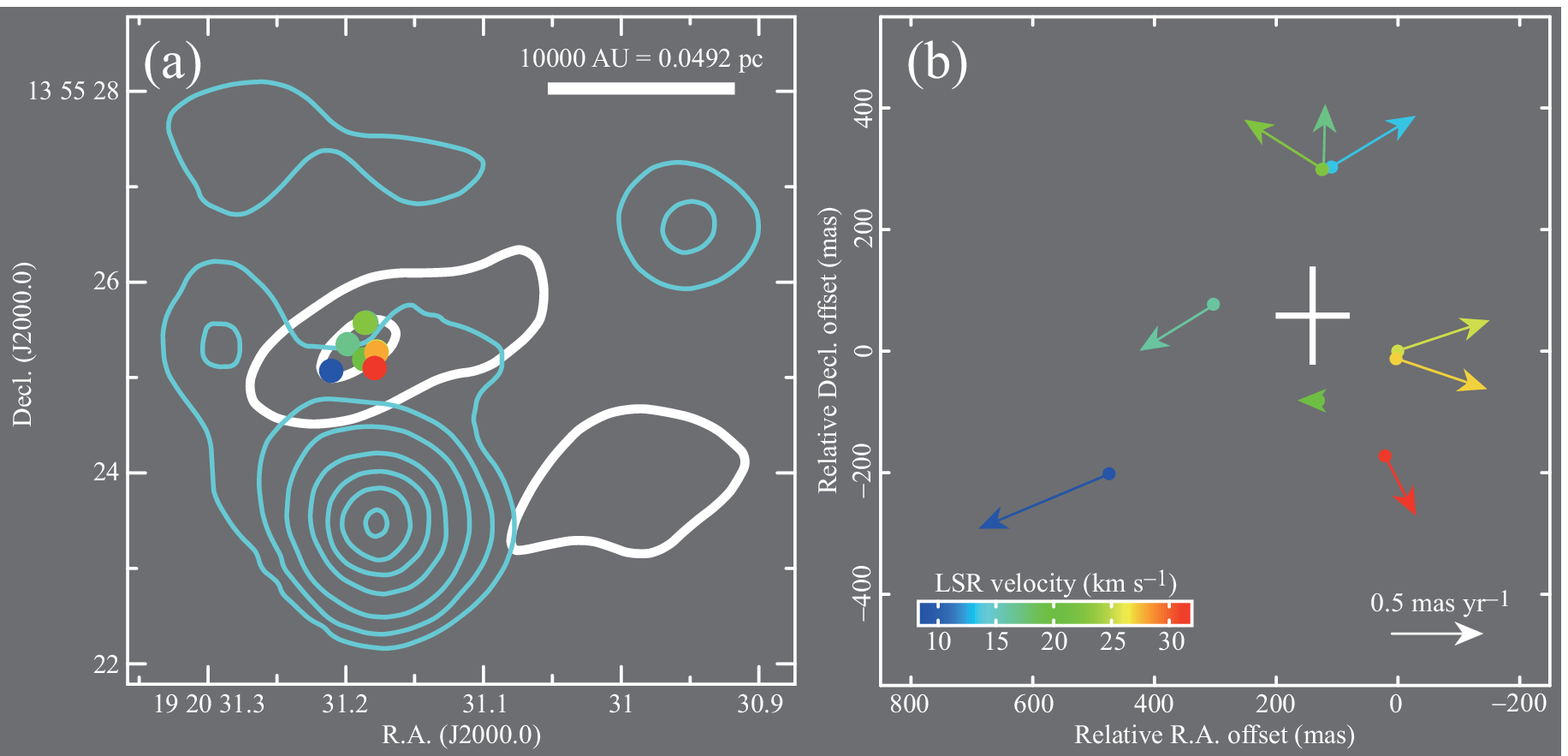}
  \end{center}
  \caption{(a): Distributions of H$_2$O masers (colord filled circle) 
                 superimposed on an 3.6 cm radio continuum map (cyan contour) (\cite{kur94})
                 The white contour shows an integrated intensity map in the NH$_3$ (2,2) line (\cite{cod10}).
            (b): Close up to the distributions of H$_2$O masers with internal-motion vectors.
                 The map origin is located at the position of the maser feature at
                 $v_{\rm LSR} = 25.5$ km s$^{-1}$, and
                 $(\alpha, \delta)_{\rm J2000.0} = (\timeform{19h20m31.17724s}, 
                 \timeform{13D55'25.2567"})$
                 in 2005/024.                 
                 The arrow at bottom-right corner indicates 
                 the internal motion of 0.5 mas yr$^{-1}$ 
                 correspoinding to 11.9 km s$^{-1}$ at the distance of 5.03 kpc.}
  \label{fig:2}
\end{figure*}

Figure \ref{fig:1} shows the scalar-averaged cross-power spectrum
of H$_2$O masers in G48.61+0.02 
observed with the Mizusawa-Iriki baseline on 2005/334.
The maser emissions were detected 
in the LSR velocity range from 8 to 32 km s$^{-1}$.
The middle of this velocity range is consistent with
the LSR velocity of the associated molecular cloud at $19\pm1$ km s$^{-1}$ 
observed in CS, C$^{18}$O, and NH$_3$ lines 
(\cite{bro96}; \cite{lop09}; \cite{cod10}).
Figure \ref{fig:2} shows the distributions of H$_2$O masers in G48.61+0.02.
The masers are distributed within approximately $500 \times 600$ mas area.
Nine maser features were detected at more than five epochs and 
seven of them were detected at all eight epochs.
They are used to measure the parallax.

\begin{figure*}
  \begin{center}
    \FigureFile(160mm,160mm){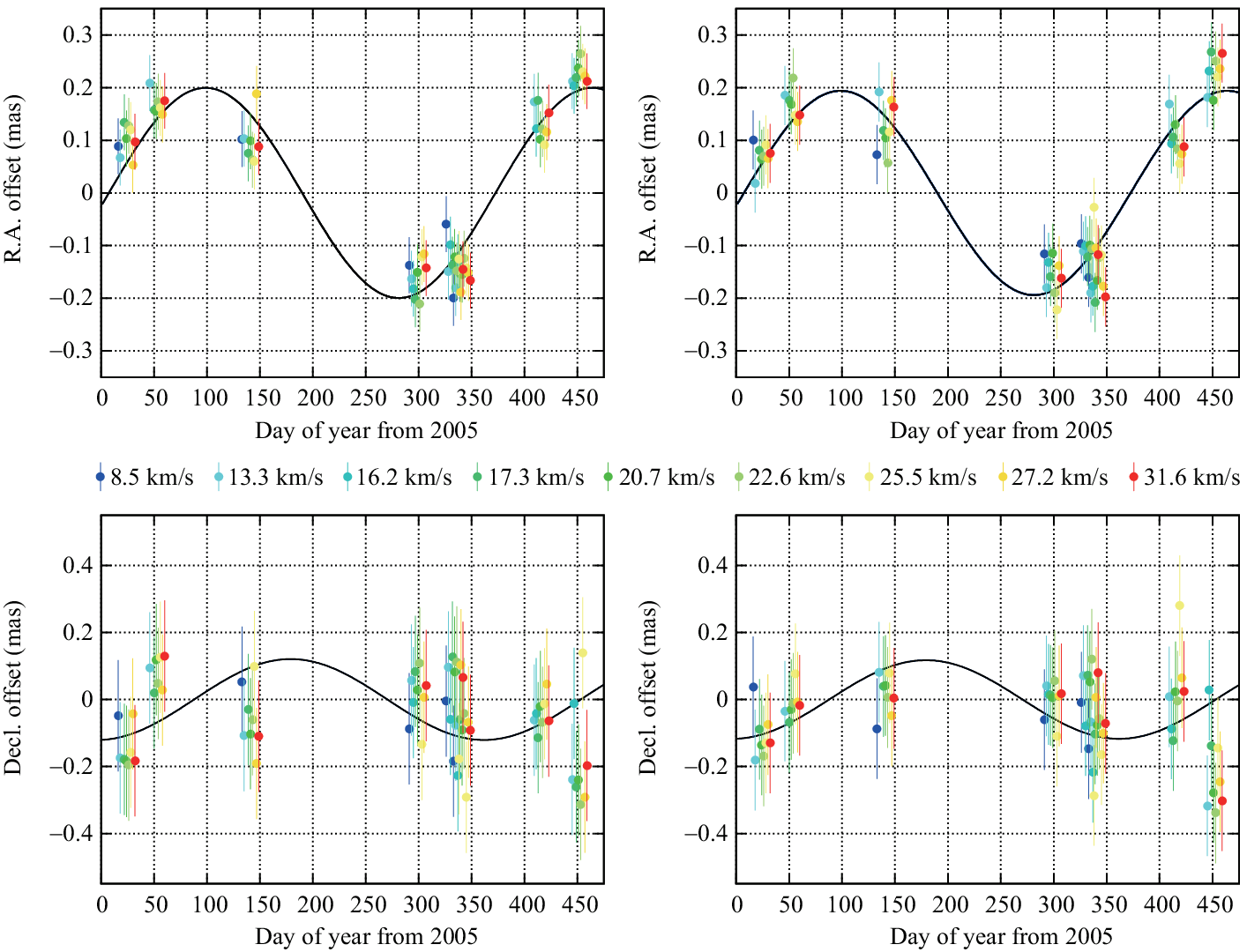}
  \end{center}
  \caption{Parallax signals for the H$_2$O masers in G48.61+0.02.
           The proper motion and the position offset are removed.
           The data for the different maser features are slightly
           shifted in time for clarity.
           The left two panels show the parallax motion obtained using J1922+1530, 
           and the right two panels show the parallax motion obtained using J1924+1540.}
  \label{fig:3}
\end{figure*}

In order to measure the parallax and proper motion of G48.61+0.02,
we conducted a combined parallax fit, 
in which the positions of nine maser features relative
to the two background sources J1922+1530 and J1924+1540 were fitted
simultaneously with one common parallax, 
but different proper motions and position offsets for individual features.
Table \ref{tab:1} and Figure \ref{fig:3} show
the results of the parallax and proper motion fit.
Since systematic errors generally dominate over random noise,
formal position errors are usually unrealistically small.
This results in relatively high reduced $\chi^2$ per degree of freedom
of parallax fit (typically values between 4 and 10).
Therefore, we assigned independent ``error floors'' in quadrature with
the formal position fitting uncertainties.
Trial combined fits were conducted and a separate reduced $\chi^2$
statistic was calculated for the right ascension and declination residuals.
The error floors of $55\mu$as in the right ascension and $158\mu$as
in declination were then adjusted iteratively so as to make the
reduced $\chi^2 \simeq 1.0$ for each coordinate.
The resulting parallax is $199\pm7$ $\mu$as.

\begin{table*}
  \caption{The measured values of parallax and proper motion 
           for H$_2$O maser features in G48.61+0.02.}
  \label{tab:1}
  \begin{center}
    \begin{tabular}{rrrccrrrrr}
      \hline
      \multicolumn{1}{c}{$v_{\rm LSR}$}                &
      \multicolumn{1}{c}{$\Delta \alpha$}             &
      \multicolumn{1}{c}{$\Delta \delta$}             &
      \multicolumn{1}{c}{Detected}                      &
      \multicolumn{1}{c}{Background}                    &
      \multicolumn{1}{c}{$\pi$}                         &
      \multicolumn{1}{c}{$\mu_{\alpha} \cos \delta$}  &
      \multicolumn{1}{c}{$\mu_{\delta}$}                &
      \multicolumn{1}{c}{$\sigma_{\alpha}$}            &
      \multicolumn{1}{c}{$\sigma_{\delta}$}            \\
      \multicolumn{1}{c}{(km s$^{-1}$)}   &
      \multicolumn{1}{c}{(mas)}           &
      \multicolumn{1}{c}{(mas)}           &
      \multicolumn{1}{c}{epochs}          &
      \multicolumn{1}{c}{source}          &
      \multicolumn{1}{c}{($\mu$as)}       &
      \multicolumn{1}{c}{(mas yr$^{-1}$)} &
      \multicolumn{1}{c}{(mas yr$^{-1}$)} &
      \multicolumn{1}{c}{($\mu$as)}      &
      \multicolumn{1}{c}{($\mu$as)}      \\
      \hline
 8.5 & 475 & $-201$ & 10111100   & J1922+1530 & ---          & $-2.08 \pm 0.10$ & $-5.58 \pm 0.11$ & 79 &  84 \\
     &     &        &            & J1924+1540 & ---          & $-1.88 \pm 0.10$ & $-5.67 \pm 0.19$ & 75 & 147 \\
     &     &        &            & Average    & ---          & $-1.98 \pm 0.07$ & $-5.61 \pm 0.10$ & 77 & 116 \\
13.3 & 109 &   303  & 11111111   & J1922+1530 & $220 \pm 26$ & $-3.20 \pm 0.04$ & $-4.95 \pm 0.16$ & 49 & 180 \\
     &     &        &            & J1924+1540 & $223 \pm 25$ & $-3.25 \pm 0.04$ & $-5.05 \pm 0.15$ & 46 & 163 \\
     &     &        &            & Average    & $222 \pm 18$ & $-3.23 \pm 0.03$ & $-5.00 \pm 0.11$ & 48 & 172 \\
16.2 & 303 &    77  & 00011111   & J1922+1530 & ---          & $-2.42 \pm 0.07$ & $-5.63 \pm 0.23$ & 24 &  79 \\
     &     &        &            & J1924+1540 & ---          & $-2.21 \pm 0.19$ & $-5.43 \pm 0.23$ & 64 &  78 \\
     &     &        &            & Average    & ---          & $-2.36 \pm 0.07$ & $-5.53 \pm 0.16$ & 44 &  79 \\
17.3 & 123 &   299  & 11111111   & J1922+1530 & $216 \pm 28$ & $-2.79 \pm 0.05$ & $-5.05 \pm 0.15$ & 51 & 173 \\
     &     &        &            & J1924+1540 & $213 \pm 30$ & $-2.75 \pm 0.05$ & $-4.94 \pm 0.09$ & 57 & 103 \\
     &     &        &            & Average    & $215 \pm 20$ & $-2.77 \pm 0.04$ & $-4.98 \pm 0.08$ & 54 & 138 \\
20.7 & 130 &  $-81$ & 11111111   & J1922+1530 & $196 \pm 24$ & $-2.71 \pm 0.04$ & $-5.23 \pm 0.15$ & 45 & 173 \\
     &     &        &            & J1924+1540 & $175 \pm 23$ & $-2.79 \pm 0.04$ & $-5.32 \pm 0.12$ & 43 & 137 \\
     &     &        &            & Average    & $185 \pm 17$ & $-2.75 \pm 0.03$ & $-5.28 \pm 0.09$ & 44 & 155 \\
22.6 & 125 &   299  & 11111111   & J1922+1530 & $218 \pm 30$ & $-2.29 \pm 0.05$ & $-5.07 \pm 0.17$ & 56 & 196 \\
     &     &        &            & J1924+1540 & $199 \pm 32$ & $-2.36 \pm 0.05$ & $-4.95 \pm 0.16$ & 60 & 182 \\
     &     &        &            & Average    & $209 \pm 22$ & $-2.33 \pm 0.04$ & $-5.01 \pm 0.12$ & 58 & 189 \\
25.5 &   0 &     0  & 11111111   & J1922+1530 & $187 \pm 31$ & $-3.27 \pm 0.05$ & $-5.15 \pm 0.12$ & 58 & 136 \\
     &     &        &            & J1924+1540 & $187 \pm 29$ & $-3.26 \pm 0.05$ & $-5.06 \pm 0.16$ & 55 & 184 \\
     &     &        &            & Average    & $187 \pm 21$ & $-3.27 \pm 0.04$ & $-5.11 \pm 0.10$ & 57 & 160 \\
27.2 &   3 &  $-12$ & 11111111   & J1922+1530 & $207 \pm 26$ & $-3.25 \pm 0.04$ & $-5.47 \pm 0.18$ & 49 & 203 \\
     &     &        &            & J1924+1540 & $193 \pm 27$ & $-3.27 \pm 0.04$ & $-5.44 \pm 0.12$ & 50 & 140 \\
     &     &        &            & Average    & $200 \pm 19$ & $-3.26 \pm 0.03$ & $-5.45 \pm 0.10$ & 50 & 172 \\
31.6 &  21 & $-172$ & 11111111   & J1922+1530 & $201 \pm 26$ & $-2.91 \pm 0.03$ & $-5.50 \pm 0.15$ & 49 & 166 \\
     &     &        &            & J1924+1540 & $216 \pm 28$ & $-2.90 \pm 0.05$ & $-5.63 \pm 0.14$ & 52 & 162 \\
     &     &        &            & Average    & $208 \pm 19$ & $-2.91 \pm 0.03$ & $-5.57 \pm 0.10$ & 51 & 164 \\
\hline
\multicolumn{4}{c}{Combined Fit} & J1922+1530 & $202 \pm 10$ &                   &                  & 53 & 166 \\
     &     &        &              & J1924+1540 & $196 \pm 11$ &                   &                  & 56 & 150 \\
     &     &        &              & All        & $199 \pm  7$ &                   &                  & 55 & 158 \\
\hline
\multicolumn{4}{c}{Average}    & J1922+1530 & $206 \pm 10$ & $-2.77 \pm 0.05$ & $-5.29 \pm 0.16$ & 51 & 154 \\
     &     &        &            & J1924+1540 & $200 \pm 10$ & $-2.74 \pm 0.07$ & $-5.28 \pm 0.15$ & 56 & 144 \\
     &     &        &            & All        & $203 \pm  7$ & $-2.76 \pm 0.04$ & $-5.28 \pm 0.11$ & 53 & 149 \\
\hline
\multicolumn{10}{@{}l@{}} {\hbox to 0pt{\parbox{170mm}{\footnotesize
Column (2), (3): Offsets relative to the positon of the maser feature at 
       $v_{\rm LSR} = 25.5$ km s$^{-1}$, and 
       $(\alpha, \delta)_{\rm J2000.0} = (\timeform{19h20m31.17724s}, \timeform{13D55'25.2567"})$ at 2005/024.\\
Column (4): `1' for detection, and `0' for non-detection.\\
Column (6): Parallax estimates.\\
Column (7), (8): Motions on the sky along the right ascension and declination. \\
Column (9), (10): Residuals of the parallax and proper motion fit.
}\hss}}
    \end{tabular}
  \end{center}
\end{table*}

To check the consistency among the parallax motions for 
individual maser feature,
we also estimated the parallax individually.
In Table \ref{tab:1}, we also show the obtained parallaxes
using individual fitting for their maser features.
We made the fitting only for seven maser features
which are detected at all eight epochs.
The 14 obtained values of the parallax using seven features and 
two background sources are consistent with each other, from 175 to 223 $\mu$as.
The error-weighted mean of these values is derived to be $203\pm7$ $\mu$as,
where the error is derived to be $\sigma^2 = 1/ \sum (1/\sigma_i^2)$.
This value is consistent with the result of the combined fit. 
We choose $199\pm7$ $\mu$as as the parallax of G48.61+0.02.
This parallax corresponds to a source distance of $5.03 \pm 0.19$ kpc.

\begin{figure}
  \begin{center}
    \FigureFile(80mm,80mm){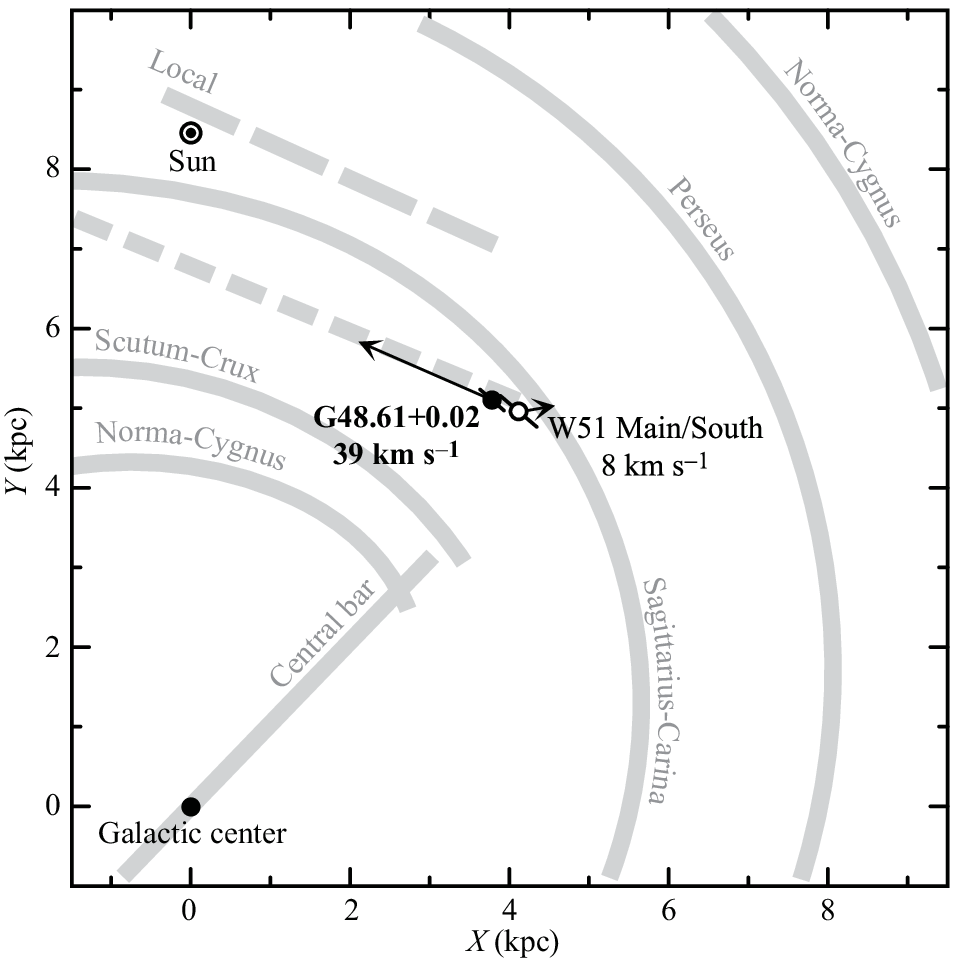}
  \end{center}
  \caption{Position of G48.61+0.02 (brack circle) superimposed on
           the four spiral arm structure of the Galaxy (\cite{rus03}), 
           and the central bar with half-length of $4.4\pm0.5$ kpc,
           tiled by $\timeform{44D} \pm \timeform{11D}$ 
           to the Sun--Galactic center line (\cite{ben05}).
           Position of W51 Main/South (white circle) 
	   observed by \citet{sat10} is also shown.
           The arrows indicates the peculiar-motion vectors
           of G48.61+0.02 and W51 Main/South which are esimated 
           in the case that $R_0 = 8.5$ kpc, $\Theta_0 = 220$ km s$^{-1}$,
           and the flat rotation are assumed.}
  \label{fig:4}
\end{figure}

The obtained distance is approximately 1/2 of the 
far kinematic distance, which was often assumed previously 
(\cite{kur94}; \cite{cod10}).
Therefore, the physical size and the luminosity of this source 
have been overestimated by factors of 2 and 4, respectively.
Figure \ref{fig:4} shows the position of G48.61+0.02 in the MWG,
which is determined from the distance of $5.03 \pm 0.19$ kpc,
the longitude of $l = \timeform{48.61D}$, and $R_0 = 8.5$ kpc.
G48.61+0.02 appears to be located on the Sagittarius-Carina arm.
We found that G48.61+0.02 is located near W51.
The obtained distance of G48.61+0.02 is very close to 
the parallactic distance of W51 Main/South, 
which is measured to be $5.41^{+0.31}_{-0.28}$ kpc 
by the H$_2$O maser astrometric observations with VLBA (\cite{sat10}).

The absolute proper motion of a maser feature is 
the sum of the internal motion of the maser feature, 
the Galactic rotation, the solar motion and the peculiar motion of the source.
All motions, expect for the internal motion, 
are common to all maser features.
Therefore, the average of the absolute proper motions of maser features
should give the systemic motion of the whole source,
in the case that the internal motion is random or symmetric.
We consider that this is valid for G48.61+0.02
because of the following two reasons.
The averaged radial velocity of all maser features is 20.3 km s$^{-1}$,
which is close to the systemic radial velocity 
derived from the associated molecular cloud.
It suggests that the internal motion is symmetric.
Figure \ref{fig:2}(b) shows the residual vectors, 
where the average of absolute proper motions is subtracted.
They should be the internal motion of maser features, and we did not find
strongly asymmetric motion.

We can divide the absolute proper motions into 
the systemic and internal motions of G48.61+0.02.
The absolute proper motions measured using two background sources 
are consistent with each other.
The averages of absolute proper motions obtained using J1922+1530 and J1924+1540
are $(\mu_\alpha \cos \delta, \mu_\delta) = (-2.77\pm0.05, -5.29\pm0.16)$ and 
$(-2.74\pm0.07, -5.28\pm0.15)$ mas yr$^{-1}$, respectively.
From the average of these values,
the systemic motion of G48.61+0.02 is derived to be
$(\mu_\alpha \cos \delta, \mu_\delta) = (-2.76\pm0.04, -5.28\pm0.11)$ mas yr$^{-1}$.
The typical value of internal motion is obtained to be 0.5 mas yr$^{-1}$,
corresponding 11.9 km s$^{-1}$ at 5.03 kpc.
This value is consistent with a half of the radial velocity span of 24 km s$^{-1}$.
The three-dimensional velocity of the internal motion is obtained to be 17 km s$^{-1}$.

\section{Discussion}

\subsection{Peculiar Motion}

G48.61+0.02 is located near W51 on the sky. 
We found that the distances of two sources are also close.
However, there is a large velocity difference between them.

The peculiar motion of G48.61+0.02 is estimated to be
$(U', V', W') = (11.2\pm1.9, -36.2\pm1.8, 6.7\pm1.5)$ km s$^{-1}$
from the measured distance of $5.03\pm0.19$ kpc and proper motion of 
$(\mu_\alpha \cos \delta, \mu_\delta) = (-2.76\pm0.04, -5.28\pm0.11)$ 
mas yr$^{-1}$, and
the radial velocity of $v_{\rm LSR} = 19 \pm 1$ km s$^{-1}$.
Here, $U'$ is the velocity component toward the Galactic Center,
$V'$ is the component in the direction of the Galactic rotation,
$W'$ is the component toward the north Galactic pole.
The peculiar motion of W51 Main/South is estimated to be
$(U', V', W') = (6.3 \pm 4.5, 5.2 \pm 4.4, 5.2 \pm 4.1)$ km s$^{-1}$
from the distance, the proper motion, and the radial velocity measured by \citet{sat10}.
In these estimations, 
we used the Galactic constants of $R_0 = 8.5$ kpc and $\Theta_0 = 220$ km s$^{-1}$ 
recommended by IAU, and assumed the flat rotation.
We also used the solar motion based on a traditional definition of 
$(U_{\odot}, V_{\odot}, W_{\odot}) = (10.3, 15.3, 7.7)$ km s$^{-1}$.
The estimated peculiar motion vectors are shown in Figure \ref{fig:4}.
In the case that we used other Galactic constants of 
$R_0 = 8.0$ kpc (\cite{rei93}) and $\Theta_0 = 236$ km s$^{-1}$,
which is consistent with measured proper motion of Sgr A* (\cite{rei04}),
the peculiar motions of G48.61+0.02 and W51 Main/South are estimated to be
$(U', V', W') = (-7.9\pm1.9, -39.4\pm1.8, 6.7\pm1.5)$ km s$^{-1}$, and
$(-16.2 \pm 4.5, 1.0 \pm 4.4, 5.2 \pm 4.1)$ km s$^{-1}$, respectively.
In both cases, 
we found that G48.61+0.02 has a large peculiar motion of $40\pm5$ km s$^{-1}$
at the three dimension,
although W51 Main/South is nearly circular orbit with no large peculiar motion.

\subsection{Origin of the Large Peculiar Motion}

As mentioned in the previous subsection,
G48.61+0.02 has the large peculiar motion of $40\pm5$ km s$^{-1}$.
The kinetic energy of the peculiar motion 
is estimated to be $(2\pm1) \times 10^{51}$ erg from 
the mass of G48.61+0.02 molecular cloud of $(1.0 \pm 0.5) \times 10^5$ \MO 
at the distance of 5.03 kpc (\cite{ohi84}).
What is the origin of this peculiar motion?

\begin{figure}
  \begin{center}
    \FigureFile(70mm,80mm){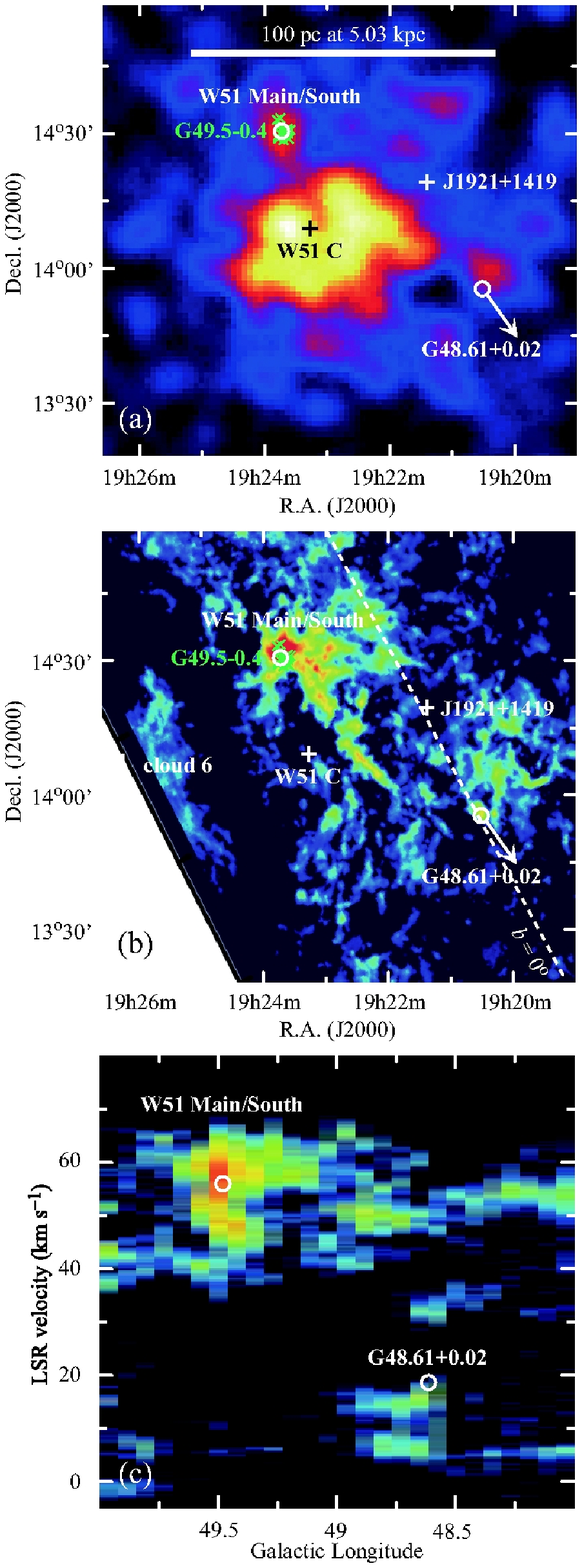}
  \end{center}
  \caption{(a): Positions of G48.61+0.02, W51 Main/South (white circle), 
            and OB stars in G49.5-0.4 (green cross)
            superimposed on $Fermi$ LAT counts map in 2--10 GeV around W51 C (\cite{gre09}).
            The crosses show the center of W51 C and the position of J1921+1419.
            The arrow shows the motion of G48.61+0.02 relative to W51 Main/South
            which is obtained to be $-0.12\pm0.16$ mas yr$^{-1}$ 
            ($-2.9\pm3.9$ km s$^{-1}$) in right ascension 
            and $-0.17\pm0.19$ mas yr$^{-1}$ ($-4.1\pm4.6$ km s$^{-1}$) in declination.
            (b): same as (a), but the background shows the integrated intensity map 
            in the $^{13}$CO $J=$ 1--0 line (\cite{jac06}).
            (c): Longitude-velocity map in the $^{13}$CO $J=$ 1--0 line (\cite{jac06}).}
  \label{fig:5}
\end{figure}

The first is the star formation activity of W51,
such as the multiple SN explosions and the radiation pressures of OB stars.
W51 C centered $(\alpha, \delta)_{\rm J2000} = (\timeform{290.818D}, \timeform{14.145D})$ 
and extended with $\timeform{0.22D}\pm\timeform{0.02}$ (\cite{abd09})
is the nearest SN remnant from G48.61+0.02.
Figure \ref{fig:5}(a) shows the position of G48.61+0.02 superimposed on
$Fermi$ LAT counts map in 2--10 GeV around W51 C.
The separation between G48.61+0.02 and the center of W51 C 
is \timeform{0.70D} (62 pc).
Figure \ref{fig:5}(b) and (c) show the integrated intensity 
and the position velocity maps 
in the $^{13}$CO $J$ = 1--0 line around W51 C (\cite{jac06}).
\citet{nak84} suggest that the molecular cloud (named ``cloud 6'' by them) 
located approximately \timeform{0.6D} east from W51 C
is physically related with the expanding shell of W51 C.
Although the large peculiar motion of G48.61+0.02 probably
cannot be driven by the single SN explosion of W51 C,
it may be originated with the multiple SN explosions in the past.
The explosion kinetic energy of W51 C is estimated to be 
$\sim 5 \times 10^{51}$ erg (\cite{abd09}).
The solid angle of G48.61+0.02 molecular cloud is 
estimated to be 0.090 sr (0.7\% of $4 \pi$ sr)
from the cloud size of \timeform{0.24D} (\cite{ohi84}) and
the separation between G48.61+0.02 and W51 C of \timeform{0.70D}.
The energy that G48.61+0.02 molecular cloud can 
receive at this solid angle from the explosion kinetic enerty of W51 C
is estimated to be $\sim 4 \times 10^{49}$ erg.
The kinetic energy of the peculiar motion requires $\sim 50$
SN explosions of the same scale as W51 C.
As the evidence of the other SN explosion,
we found that a pulsar J1921+1419 (\cite{man05})
is located at \timeform{0.45D} (40 pc) northeast from G48.61+0.02.

W51 is a rich birth site of OB stars.
Their radiation pressures also may be possible origin of the large peculiar motion.
The most luminous source in W51 is 
the G49.5-0.4 H\emissiontype{II} region complex
which is located at \timeform{0.97D} (85 pc) northeast from G48.61+0.02 
(see Figure \ref{fig:5}(a)).
There are 34 O stars and hundreds of B stars in G49.5-0.4 (\cite{oku00}).
Their total bolometric luminosity and the age are estimated to be 
approximately $9 \times 10^6 \LO$ at 5.03 kpc and 1 Myr, respectively (\cite{oku00}).
From these values, the energy that have been radiated from the OB stars
since they were formed is estimated to be $1 \times 10^{54}$ erg.
The energy of the radiation that G48.61+0.02 cloud 
can receive at this solid angle
is estimated to be $4 \times 10^{51}$ erg 
from the solid angle of 0.048 sr (0.4\% of $4 \pi$ sr) estimated from 
the separation between G48.61+0.02 and G49.5-0.4 of \timeform{0.97D}
and the cloud size of \timeform{0.24D} (\cite{ohi84}).
The kinetic energy of the peculiar motion is approximately 50\% of this energy.
Because as much as half of the radiative energy would be efficiently 
transformed into the kinetic energy,
the large peculiar motion would not be driven by the radiation pressures.
Considering the energetics, the multiple SN explosions are the possible origin.

The second is the gravitational potential of the central bar.
The large noncircular motions expected to be originated by the central bar
are observed in the inner Galactic star-forming regions,
G23.01-0.41, G23.44-0.18, and G23.657-00.127 (\cite{bar08}; \cite{bru09}).
The peculiar motion velocity of G48.61+0.02 is close to 
those of three star-forming regions.
However, G48.61+0.02 is approximately 2 kpc away from the end of the central bar 
(see Figure \ref{fig:4}), and it is further than three regions from the central bar.
In addition, the similar large peculiar motion is
not found in the nearby source W51 Main/South.
Therefore, we suggest that 
the peculiar motion of G48.61+0.02 is originated with 
the local structure of less than the hundreds pc,
rather than the central bar with the kpc scale.

G48.61+0.02 appears to be located on the Sagittarius-Carina arm.
Therefore, the streaming motion across the spiral arm
is one possible origin for the large peculiar motion of G48.61+0.02.

\subsection{Driving Source of H$_2$O Masers}

The internal motions exhibit an expanding outflow structure.
We could obtain the dynamical center of the outflow
to be $(\Delta \alpha, \Delta \delta) = (140\pm60, 60\pm80)$ mas,
based on a simple model that
the maser features were projected with the
measured internal motions from the single origin at the same time.
This center is shown by a cross symbol in Figure \ref{fig:2}(b).

To investigate the presence of driving source of H$_2$O masers,
we compare the maps observed with the high angular resolutions.
The 3.6 cm radio continuum emissions
and the NH$_3$ $(J,K) = (2,2)$ emissions observed with the VLA 
(\cite{kur94}; \cite{cod10})
are shown in Figure \ref{fig:2}(a).
The H$_2$O masers are located at approximately 
\timeform{2"} (10000 AU at 5.03 kpc) north
from the peak position of the 3.6 cm continuum source, G48.606+0.023,
and are spatially coincident with the NH$_3$ core.

The H$_2$O masers in G48.61+0.02 
could be associated with a massive young stellar object (YSO) 
in the NH$_3$ core.
\citet{cod10} estimate the virial mass of this NH$_3$ core
to be 337 \MO at their assumed distance of 9.7 kpc.
This mass is re-estimated to be 173 \MO at the distance of 5.03 kpc.
In the case that we assume 
the typical star formation efficiency of $\simeq$ 10--30\% (\cite{lad03}),
the stellar mass of YSOs formed in the NH$_3$ core is estimated to be $\simeq$ 17--52 \MO.
The bright infrared emission corresponding to this mass is detected in this region.
The infrared luminosity is estimated to be  $1.7 \times 10^5 \LO$ 
at the distance of 5.03 kpc (\cite{ver03}).


\section{Conclusions}

We observed the H$_2$O masers in the Galactic star-forming region G48.61+0.02 with VERA.
Our conclusions are summarized as follows:

\begin{enumerate}

  \item 
We have determined the trigonometric parallax of G48.61+0.02 to be 
$199 \pm 7$ $\mu$as, corresponding to a distance of $5.03 \pm 0.19$ kpc.
This distance places G48.61+0.02 in the Sagittarius-Carina arm and near
the active star forming region and the SN remnant W51.

  \item
The observed distance, proper motion, and radial velocity
of G48.61+0.02 indicate that 
it has a large peculiar motion of $40\pm 5$ km s$^{-1}$.
This large peculiar motion might be the result of the multiple SN explosions
in W51, or the streaming motion across the Sagittarius-Carina arm.

  \item
The internal motions of H$_2$O masers in G48.61+0.02 
exhibit the expanding outflow with the three-dimensional velocity of 17 km s$^{-1}$.
The masers would be associated with the YSO 
with the mass of approximately 17--52 \MO formed in the NH$_3$ core.

\end{enumerate}

\bigskip

We are grateful to an anonymous referee for valuable comments and suggestions.
We thank to the staff members of all the VERA stations
for their assistances in the observations.


\end{document}